\documentclass[]{spie}  

\usepackage{amsmath,amsfonts,amssymb}
\usepackage{graphicx}
\usepackage[colorlinks=true, allcolors=blue]{hyperref}

\title{Polarization dOTF: on-sky focal plane wavefront sensing}

\author[a]{Keira J. Brooks}
\author[b,c]{Laure Catala}
\author[d]{Matthew A. Kenworthy}
\author[b]{Steven M. Crawford}
\author[e]{Johanan L. Codona}
\affil[a]{Space Telescope Science Institute, 3700 San Martin Dr, Baltimore, MD, 21218 USA}
\affil[b]{South African Astronomical Observatory, Observatory Rd., Cape Town, South Africa}
\affil[c]{Department of Astronomy, University of Cape Town, Private Bag X3, Rondebosch 7701, South Africa}
\affil[d]{Leiden Observatory, Leiden University, P.O. Box 9513, 2300, RA Leiden, The Netherlands}
\affil[e]{Steward Observatory, University of Arizona, 933 N. Cherry Ave, Tucson, AZ USA 85721}

\authorinfo{E-mail: kbrooks@stsci.edu}

\begin{document}
\maketitle

\begin{abstract}

The differential Optical Transfer Function (dOTF) is a focal plane wavefront
sensing method that uses a diversity in the pupil plane to generate two
different focal plane images.
The difference of their Fourier transforms recovers the complex
amplitude of the pupil down to the spatial scale of the diversity.
We produce two simultaneous PSF images with diversity using a polarizing
filter at the edge of the telescope pupil, and a polarization camera to
simultaneously record the two images.
Here we present the first on-sky demonstration of polarization dOTF at
the 1.0m South African Astronomical Observatory telescope in Sutherland,
and our attempt to validate it with simultaneous Shack-Hartmann
wavefront sensor images.

\end{abstract}


\section{INTRODUCTION}
\label{sec:intro}  

One of the greatest challenges with ground-based observations is the
perturbation of the incoming wavefronts through the turbulent atmosphere.
We approximate light from a distant object as a succession of plane
parallel flat wavefronts, but by the time the starlight reaches the ground,
it is no longer flat.
If we are able to measure the atmospheric aberrations through wavefront
sensing, it is possible to correct for them using an adaptive optics
system or through post-processing of the point spread function (PSF).

The Optical Transfer Function (OTF) can be used to measure the image
quality of an optical system, defined as the Fourier transform of the
amplitude squared of a point source, to that of the image in frequency
space (\cite{2002Williams}).
Using the definition of the OTF, \cite{2012Codona, 2013Codona} it was
shown\cite{2012Doble} that taking the difference of the OTFs of two
pupils with an amplitude or phase perturbation can return information
about the identical regions of the pupils, notably the complex amplitude
of the electrical field in the pupil.
In order to reconstruct this phase, the two images of the complex pupils
must be taken in quick succession, during which time the phase must be
expected not to change so that the only difference measured is the
modification made to the pupil.
Given the rapid temporal evolution of the atmosphere, it is impractical
to image two different pupils so quickly as to image the phase of the
atmospheric wavefront before it changes, without also being affected by
any vibrations in the system.

Using polarization for dOTF was first explored using a polarization
finger in the pupil plane and a polarized beamsplitter to split the
light into two different polarization orientations \cite{2012Codona}.
This method, however, was rejected due to the introduction of both fixed
and temporal non-common-path errors that are impractical to remove.
Instead, we use a camera with a pixel-matched micropolarizer array to
perform dOTF without any common path errors and without the need for a
centrated star.
We carry out first on-sky observations with a polarization camera and a
Shack-Hartmann sensor to validate the reconstructed atmospheric phase.
This method, which we call polarization dOTF, is still limited by PSF saturation and
signal-to-noise on the detector.

We show through on-sky testing that this method produces a measure of
diversity and we analyze the data we have obtained.
Due to camera timing issues, we were not able to validate the
polarization dOTF method with the Shack-Hartmann wavefront sensor.

\section{Polarization dOTF}

\label{sec:pol_dotf}

\begin{figure}
	\centering
	\includegraphics[width=.7\linewidth]{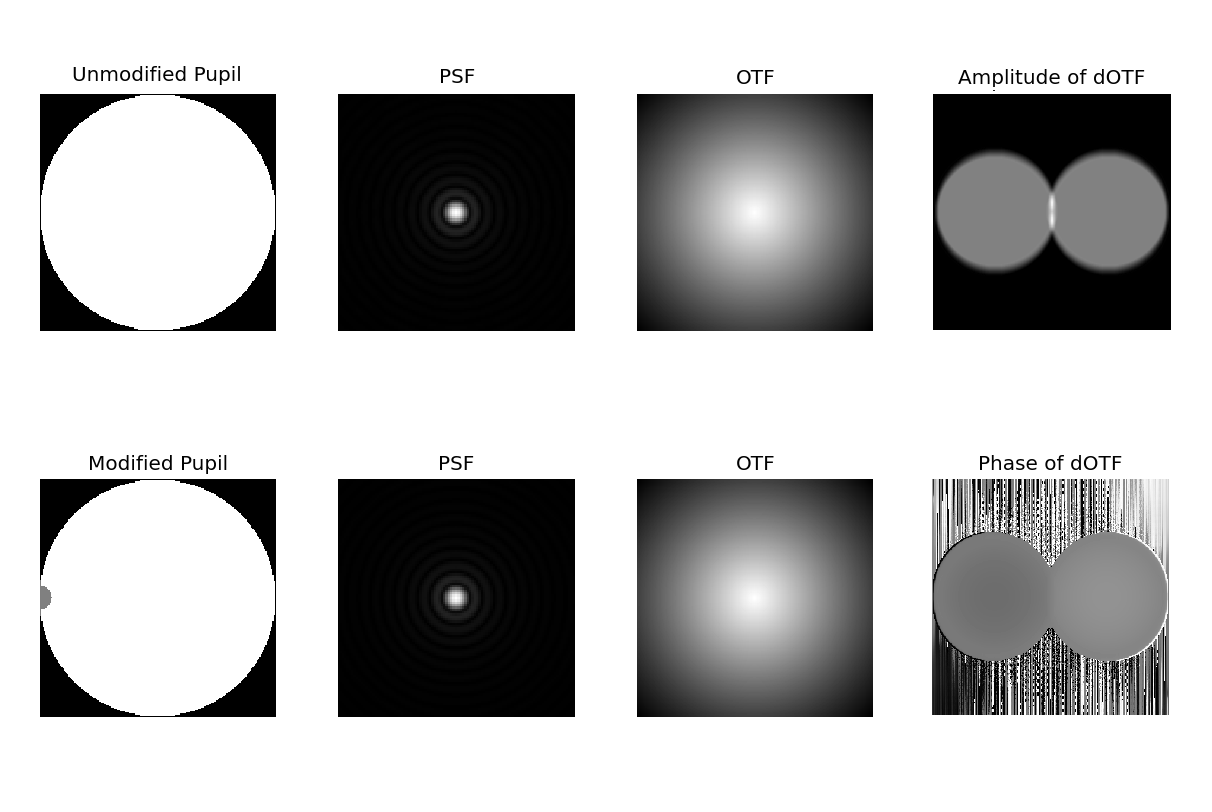} 
    \caption{Explanation of the dOTF method. We use two different pupil
planes (leftmost column) to image the PSF of a star (second column). We
take the Fourier transform of those PSFs, giving the OTF (third column).
The difference of the two OTFs gives the amplitude and phase of the
pupil plane, convolved with the perturbing function (fourth column).}
    \label{fig:sample_dotf}
\end{figure}

The image-based differential Optical Transfer Function wavefront sensing
(dOTF WFS) method was first introduced \cite{2012Codona, 2012Doble,
2013Codona} as a supplement to current wavefront sensing methods. 
This method requires the imaging of two similar pupils, where one pupil
is modified.
The Fourier transform of each PSF produced by the pupils give their
respective OTFs, which are then subtracted from one another. 
The resultant image is of two overlapping, complex conjugates of the
initial pupils, reflected about the area where they are different (see
Figure~\ref{fig:sample_dotf}).
Any part of these pupils not in the overlap region is the complex
conjugate of the original complex pupil plane\cite{2013Codona} which is
a consequence of taking the Fourier transform of the PSF.
This unique property allows us to split the dOTF image into both
amplitude and phase, where the phase is the phase of the wavefront at
the time of observation. 
To retain the maximum amount of information about the original amplitude
and phase, the modification is placed at the edge of the
pupil, minimizing the loss of information in the overlap region.

While all previous iterations of dOTF WFS (where images are taken
in quick succession) provide good diagnostics on the method, in order
to be sure that we are imaging the same phase of the wavefront for each
pupil, simultaneous imaging needs to occur, something that can be
achieved with polarization dOTF.

\subsubsection{Using polarization}

Light is a transverse electromagnetic wave that can be decomposed into
two orthogonal polarizations. 
The direction or angle of polarization is often described by the Stokes
vector [I,Q,U,V]$^{T}$.
Here we focus only on the linear polarization: Q,
the horizontally and vertically polarized light, and U, the diagonally
polarized light. 

We explore a detector that can register two orthogonal directions of
linear polarization simultaneously (+Q and -Q, or +U and -U). 
By using a linearly polarized modification in the pupil plane that is
aligned with one of the measurable directions, we can retrieve two
images - one for each polarization orientation - given that two
orthogonally orientated polarizing sheets block out all light, and
parallel orientated polarizing sheets allows all light to pass. 
While this removes many of the previously presented challenges with the
dOTF method for WFS, polarization dOTF WFS is still limited by PSF
saturation and low signal to noise.

\section{Observations}

During 2015 June 03 to 09 UT, we observed at the South African
Astronomical Observatory (SAAO) in Sutherland, South Africa.
During these six nights of observation, we tested our Polarization
Camera alongside a Shack-Hartmann Wavefront Sensor (SHWFS) within a
custom built cage mount. 

\subsection{Site}

The SAAO Sutherland site is located at an altitude of 1798m,
approximately 375km outside of Cape Town in South Africa. This site is
home to the Southern Africa Large Telescope (SALT) and several other
telescopes of varying size.
The optical turbulence at the Sutherland site \cite{2013Catala} from the
surface to an altitude of 1km above the surface at this site has a
median seeing of 1.32'', accounting for 84\% of the turbulence. 
The other 16\% comes from the atmosphere above 1km with a median seeing
of 0.41''.

\subsection{Instrument Setup}

\begin{figure}
	\includegraphics[width=\columnwidth]{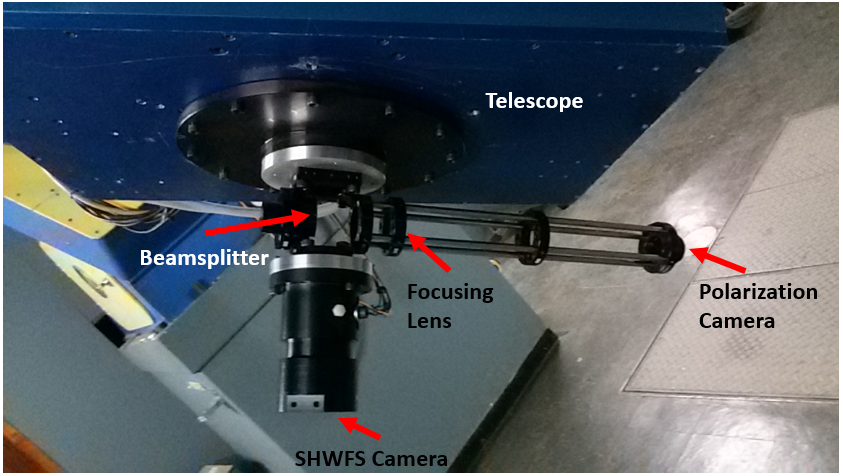}
    \caption{The cage for the polarization camera and support for the
SHWFS on the 1m telescope at SAAO, Sutherland site. This image is shown
without optical components for clarity.}
    \label{fig:tel_setup1}
\end{figure}

At the SAAO Sutherland site, we used the 1m Cassegrain reflector, adding
our instrument to the back of the telescope as seen in
Figure~\ref{fig:tel_setup1}.
In order to simultaneously measure the wavefront with the SHWFS and the
polarization camera, we used a 50\% beamsplitter to feed light to both
wavefront sensors simultaneously.

Given simultaneous imaging on two detectors, the instrument included a
beamsplitter that sent half of the light to the SHWFS (mounted directly
behind the telescope) and the other half to the polarization camera. 
In the pupil of the telescope we taped a circular polarization sheet,
10cm in diameter, to act as our obscuration, and aligned it with respect
to one of the pixels in the polarization camera.

The camera used for these tests is a 4DTechnology PolarCam$^{TM}$,
pixelated polarizer-based CCD\cite{2011Brock}. 
This camera is unique in its ability to image in four different
polarization angles simultaneously. 
This is possible through the use of a pixelated micropolarizer array
which is matched to the pixels on the detector. 
This polarizer array is made up of many ``super pixels'', where one
super pixel includes four pixels, each with a different linear
polarization orientation (+Q, -Q, +U, -U).
By mounting the micropolarizer array directly onto the CCD and adding a
border around each polarizing element, polarization crosstalk is
limited.
At 550nm, the maximum transmission of the pixelated polarizer is 80\% of
the total amount of light that is polarized in that
direction\cite{2011Brock}. 
See Table~\ref{tab:polcam_specs} for additional specifications for this
camera.
An optical cage system was used to mount the optical components for the
polarization dOTF WFS, including the beamsplitter, a focusing lens, and
the polarization camera, attached to the end of the optics cage (seen to
the right in Figure~\ref{fig:tel_setup1}).

\begin{table}
	\centering
	\caption{Specifications of the 4DTechnology PolarCam$^{TM}$ Polarization Camera. (From the data sheet)}
	\label{tab:polcam_specs}
	\begin{tabular}{ll} 
		\hline
		Description & PolarCam Polarization Camera\\
		\hline
		Acquisition Mode& Simultaneous polarization imaging	\\
        Sensor Type		& Monochrome interline transfer CCD\\
		Spectral Range	& 320 nm - 1060 nm\\
		Pixel Size		& 7.4 $\mu$m	\\
        Bit Depth		& 12-bit		\\
        Interface		& GigE Ethernet	\\
        Camera Speed	& 114 fps (maximum)\\
		\hline
	\end{tabular}
\end{table}

\begin{figure}
	\centering
	\includegraphics[width=.5\columnwidth]{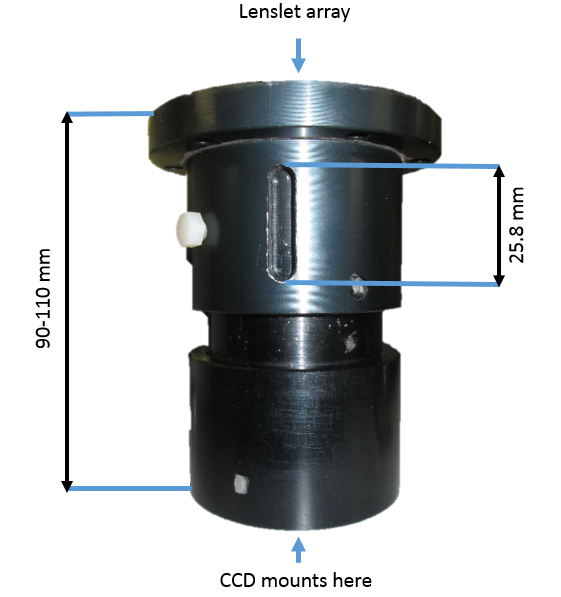}

    \caption{Photograph of SHWFS system without the detector.
This system includes a collimating lens at the top, and then a nested
cylinder to which a camera can be attached.
The slit in the side is where a screw is put to secure the inner
cylinder and is used to find the focus.
This slit is 25.8mm long, however leaving room for the screws, the focus
can only be adjusted by about 20mm.}

    \label{fig:SHWFS_schematic}
\end{figure}

The SHWFS used for observation was assembled such that the light was
first sent through a collimating lens and then to the camera on which
there is a lenslet array forming a Shack-Hartmann WFS.
A schematic of this system can be seen in
Figure~\ref{fig:SHWFS_schematic}, showing that the collimating lens sits
near the top of the system (in the figure), below which there is a
nested cylinder that can be extended by up to 20mm.
The detector is mounted to the bottom of this system which was then attached
below the beamsplitter, as seen in Figure~\ref{fig:tel_setup1}.
 
\subsubsection{Polarization Obscuration Alignment}

In order for polarization dOTF to work properly, the obscuration at the
telescope pupil must have its angle of polarization aligned with one of
the four polarization angles in each super pixel of the camera.
If the polarization angle of the camera is, for example, aligned with
the  +Q and -Q orientations, then we can use those pixels on the
detector.
The same goes for the +U and -U orientations.

\begin{figure}[!htb]
\centering
\minipage{0.32\textwidth}
\includegraphics[width=\textwidth]{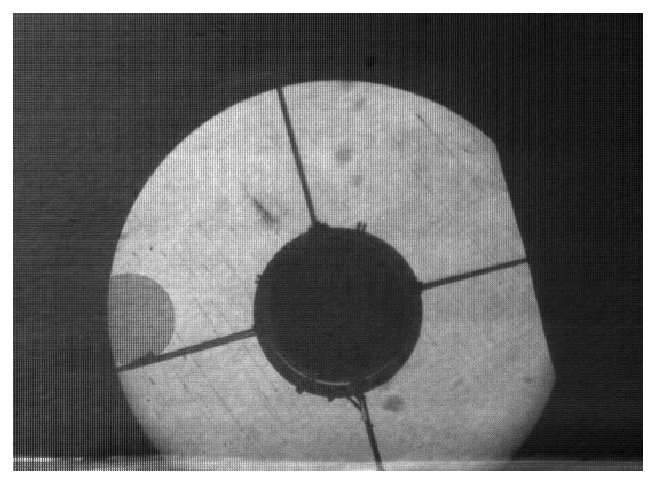}
\endminipage\hfill
\minipage{0.32\textwidth}
\includegraphics[width=\textwidth]{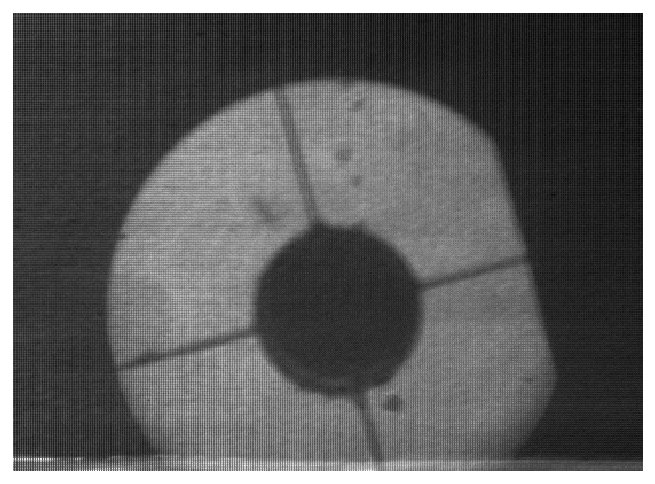}
\endminipage\hfill
\minipage{0.32\textwidth}%
\includegraphics[width=\textwidth]{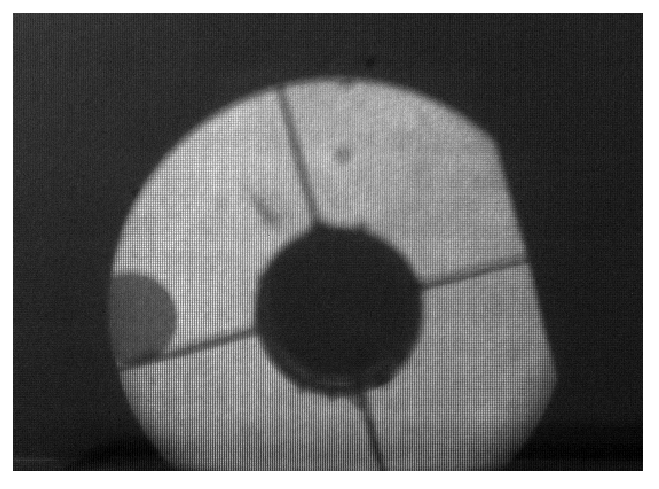}
\endminipage

\caption{The pupil images taken while performing the checks on the
direction of polarization.
The second image shows when the small piece of polarizer is placed
parallel to the orientation of the obscuration in the pupil plane.
The third image shows the opposite; when the piece of polarizer is
orientated perpendicular to the polarization orientation of the
obscuration in the pupil plane.
These images have not been split into their polarization orientations.} 

\label{fig:poltest_finger}
\end{figure}

Imaging the pupil plane with the polarization finger in position and
aligned with the camera, we see that $\approx 50\%$ of the light is lost
at the location of this obscuration (see
Figure~\ref{fig:poltest_finger}, left).
In order to align the camera and the polarization finger, a secondary
polarization finger was made that could be placed directly in front of
the camera and rotated until uniform intensity is reached (half the
intensity of the original image, middle image) across the pupil plane,
or zero throughput (Figure~\ref{fig:poltest_finger}, right) at the point
at which the polarization obscuration sits in the pupil plane.

\subsection{Observation Overview}

Due to the setup of our system, splitting the light between the SHWFS
and the polarization camera, it was necessary to observe stars brighter
than 2\textsuperscript{nd} magnitude.
In a rough estimate of the loss of light through the elements in the
telescope and our instrument (as outline in Table~\ref{tab:throughput}),
we can only expect a throughput of $\approx 3.5 \%$. Additionally, we are limited in our selection of objects to single-star systems, since the lenslets of the SHWFS make it difficult to
extract phase information from a binary system unless the stars have a sufficiently large separation. An exception for one binary system (Antares) was made due to brightness limitations.

\begin{table}
\centering
\caption{Estimated throughput of the instrument}
\label{tab:throughput}
\begin{tabular}{lc}
\hline
\hline
Optical Component                   &   Transmission     \\
\hline
Atmosphere									&  80\%\\
Primary Mirror               				&   60\%  \\
Secondary Mirror						&  60\% \\
Beamsplitter									&    50\%  \\
Detector Quantum Efficiency &   60\% \\
Polarization									& 	40\%		\\
\hline
Percent of light remaining 					&  3.456\% \\
\end{tabular}
\end{table}

We have two nights of usable observations, during which we observed several bright objects with exposure times between 10 and 30ms to try to match the timescale on which the atmosphere changes. Simultaneously, we collected seeing measurements from the Multi Aperture Scintillation Sensor - Differential Imaging Motion Monitor (MASS-DIMM) device at the Sutherland site. The seeing varied from 1.0-1.2'' on 7 June and 1.4-1.5'' on 8 June. 

\section{Analysis}

\subsection{Modelling: What we expect}
\label{3simulations}

To get an idea of what to expect, we simulate dOTF wavefront sensing
based on the conditions of the 1m telescope at the SAAO Sutherland site.

For our simulations, we use a pupil mask that mimics the pupil of the 1m telescope with a secondary mirror supported by four beams.
To the pupil we add an obscuration of varying sizes, at 50\%
intensity to imitate a polarized obscuration.

\begin{figure}
	\centering
    \includegraphics[width=.75\linewidth]{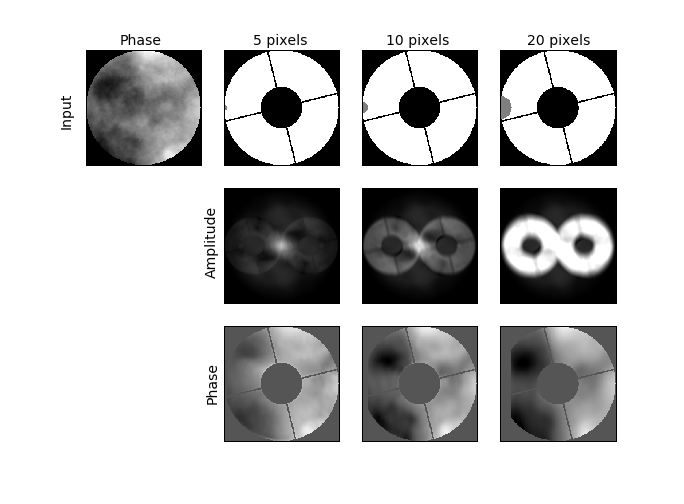}
    \caption{Three different obscuration radii (5, 10, and 20 pixels)
show how changing the pupil can affect the amplitude and phase of the dOTF outcome using a Kolmogorov phase screen. We introduce a Gaussian noise distribution to the PSF and reduce the flux by a factor of 10$^{3}$ to simulate realistic conditions. At the top of the figure, we show the different obscuration sizes, increasing from the left. In the second row, we show what we expect for the amplitude of the dOTF given the input phase and noise additions. Directly below each amplitude image, we show the phase of the reconstructed wavefront, masked for clarity. The scale for the reconstructed amplitude is 0 to 750,000 counts, while the scale for the reconstructed phase is -$\pi$ to $\pi$ radians.}

    \label{fig:obs_kol}
\end{figure}

Choosing the optimum size of the obscuration is based on the balance
between the effects of photon shot noise, integration time, and spatial
resolution.
A small obscuration will return more of the higher order structure of the phase in the pupil plane,
however, it requires a longer integration time in order to get enough
signal to overpower the photon shot noise, and integration time should
not exceed the coherence time of the atmosphere in order to properly image the changing wavefronts.
A large obscuration can result in a loss of spatial resolution since the
variations in the phase become smeared out.
Additionally, the size of the overlap region corresponds to the size of
the obscuration disk and in this region, the phase map will not be reconstructed, resulting in an
even greater loss in phase information for larger obscurations.
Through our simulations, we can explore this balance in parameter space (as done in Figure~\ref{fig:obs_kol}) and compare this to the results of our on-sky testing.

In the upper left corner of the top figure in Figure~\ref{fig:obs_kol},
we show the phase applied to the initial complex pupil plane. Here we use a Kolmogorov phase screen, based on the
Kolmogorov model of atmospheric turbulence, which seeks to model the
random motions of atmospheric turbulence by looking at motion in a fluid
medium\cite{1998Hardy}.
We also add a Gaussian noise distribution to the PSF and reduce the PSF
flux by 1,000 in order to model a non-ideal case.
To the right of the phase map in Figure~\ref{fig:obs_kol}, we show the
three different obscuration sizes in the expected telescope pupil - radius = 5, 10, and 20 pixels, for an aperture of D = 200 pixels - where half of the obscuration can be seen on the left side of the pupil.
For each combination of telescope pupil and obscuration size, using the
indicated phase map, we find the dOTF reconstructed amplitude(second row) and phase(third row).

As was indicated in Figure~\ref{fig:sample_dotf}, the dOTF reconstructed phase
includes a large amount of noise as a result of the Fourier transform,
so as to clearly seen how the dOTF reconstructs the phase, we mask
one of the circles and orient it to match the input phase.
We also mask the secondary, secondary supports, and the overlap
region in the pupil plane as there is no information about the phase retained in these regions.

In the amplitude of the dOTF image (Figure~\ref{fig:obs_kol}, second row) the
simulations show us that in the presence of noise, a one pixel
obscuration will result in a dOTF reconstructed image dominated by
noise, and we easily see that there is much more diversity with the
large obscuration.
In the masked phase (Figure~\ref{fig:obs_kol}, third row) we see that the phase is poorly 
reconstructed with small obscurations, while the use of
larger obscurations tends toward better reconstruction results.

\begin{figure}
	\centering
	\includegraphics[width=.75\linewidth]{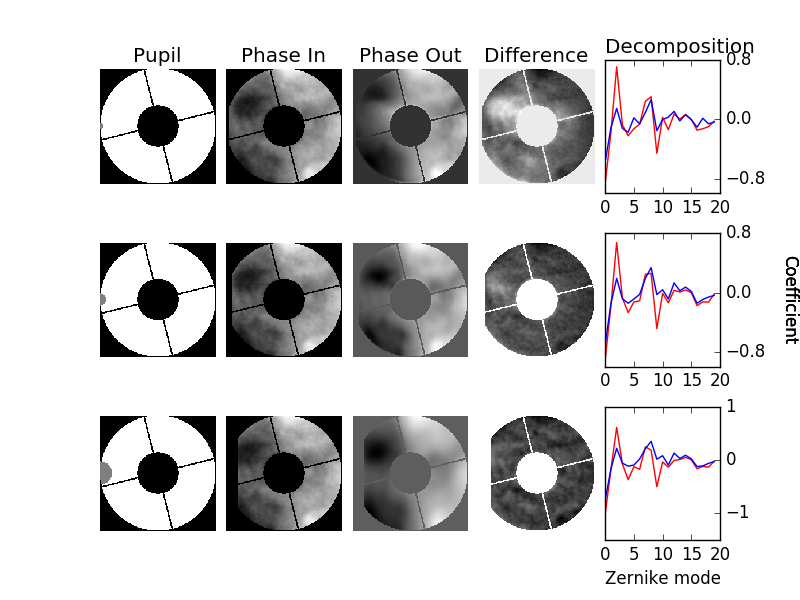}
    \caption{Using the same input phase map as the one indicated in
Figure~\ref{fig:obs_kol}, we show the input pupil used in the first
column (with obscuration radii of 5, 10 and 20 pixels, respectively),
the second column gives the masked input phase, the third column shows
the the masked output phase, and the fourth column gives the difference
between the two. The last column shows the decomposition of the input
and output phase maps in blue and red, respectively. The decomposition
shows the expected Zernike coefficient necessary for each mode in order to reconstruct the phase maps. The input phase scales from 0 to 2$\pi$ radians and the scale for the reconstructed phase is -$\pi$ to $\pi$ radians.}
    \label{fig:zernike_decomp3}
\end{figure}

Moving one step forward to see how well we can reconstruct the phase, we turn to the
Zernike decomposition of our input and output (reconstructed) phase
given the convolution with the pupil. This can tell us which Zernike modes are most difficult for this method of wavefront sensing to reconstruct 
Figure~\ref{fig:zernike_decomp3} shows the input pupil with the
different sizes of obscurations, the input phase masked with the pupil, the
dOTF reconstructed phase, the difference between the second and third
columns, and the decomposition of the second (in blue) and third (in
red) columns into their Zernike coefficients.
The last column clearly shows where the reconstruction of the phase
struggles (the regions where the red and blue lines diverge). The modes where such divergence is prominent, correspond with regions where the phase is masked (i.e. the center of the reconstruction). As a result, we maintain that we expect the regions where the pupil remains unmasked, the dOTF reconstruction of the phase, even for larger obscurations, is able to retain a large amount of the phase structure.

\section{Results}
\begin{figure}
	\centering
	\includegraphics[width=\linewidth]{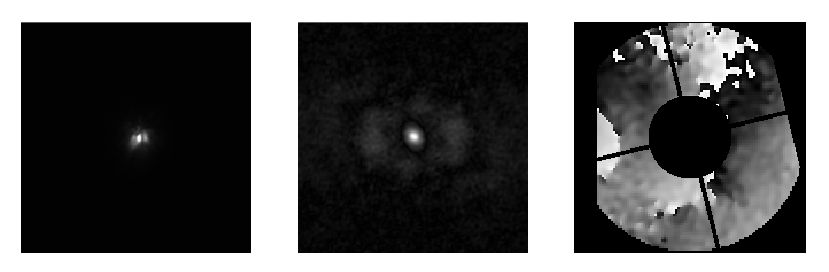}
    \caption{The star Antares, observed with no filter for an exposure time of 30ms is shown in the first image on the left. The middle and right images are the results of dOTF in the amplitude (middle) and masked phase (right). We see that the resultant amplitude image is most likely noise dominated meaning that the phase reconstruction image is also likely to be dominated by noise. }
    \label{fig:dOTF_final_frame}
\end{figure}

\begin{figure}
	\centering
	\includegraphics[width=\linewidth]{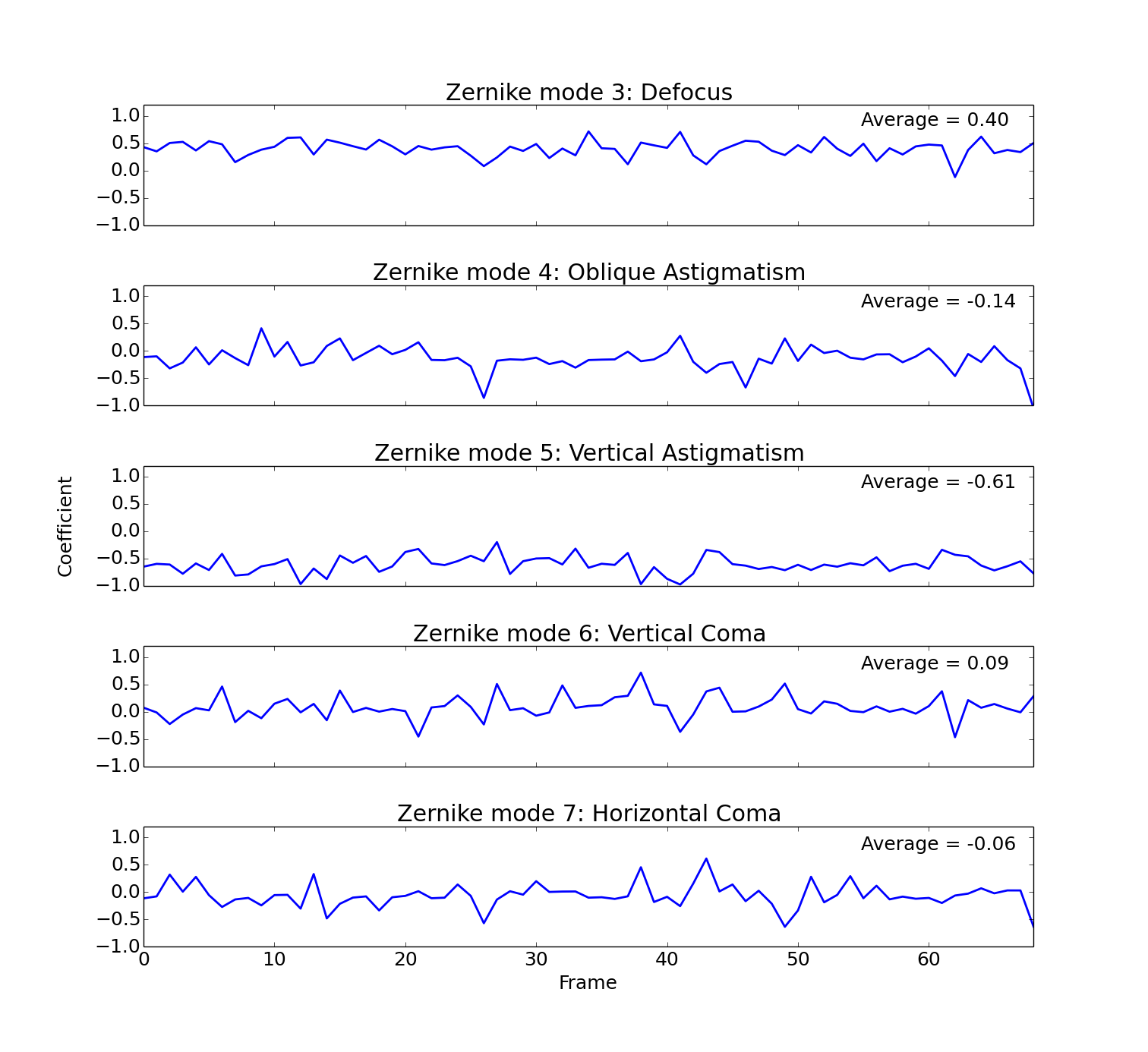}
    \caption{Zernike decomposition for all 69 frames of Antares (30ms exposures and no filter). This includes tilt, tip, defocus, astigmatism, and coma aberrations. The mean coefficient values for each polynomials is given. Due to the lack of coherent trends, we conclude that our observations are dominated by noise.}
    \label{fig:dOTF_z_decomp} 
\end{figure}

With the reduced data from the observations split into the different polarization orientations,
we are able to perform dOTF on the data for the +Q and -Q orientations.
We can expect to recreate the phase of the wavefront at that moment and
create a movie of how the phase of the wavefront changes during the
moments of observation.
We expect that the variety of stars, exposure times, and filters, will
give us limits on the dOTF method.
Unfortunately, instead we find that for most observations, there is not
enough signal from the star to fully reconstruct the phase.
We show in Figure~\ref{fig:dOTF_final_frame} one of the exposures of
Antares with an exposure time of 30ms and no filter, with the amplitude,
and the masked phase the dOTF reconstruction. 
The reconstruction appears to be heavily influenced by noise and we do not see the expected diversity given the size of the obscuration. Even when looking through a movie of the reconstructed amplitude and phase for all frames during the observation of Antares, we see a change in the signal and reconstructed phase, however patterns moving across the pupil plane do not emerge as expected. We attempt to further discern these patterns by breaking down each frame into several Zernike modes (as seen in Figure~\ref{fig:dOTF_z_decomp}) and seeing how the coefficients of those modes evolve with time. Smooth trends would emerge if the signal were not dominated by noise, but as is clearly seen in Figure~\ref{fig:dOTF_z_decomp}, there are no trends to follow. 

Using the Kolmogorov phase screen (Figure~\ref{fig:obs_kol}), we see
that by adding noise and reducing the flux, for obscuration radii of 1
or 5 pixels, the dOTF reconstructed amplitude was dominated by noise and
takes on a similar shape to our dOTF reconstructed amplitude data.
With obscuration radii on the same order as the one used on-sky (1/10 of
the aperture), the simulations showed more signal in the output.

\subsection{Camera Performance}

We found that the micropolarizer array was seamless with no measured
polarization crosstalk.
Additionally, we had no problems with the polarization of the light.
The light lost due to polarization does play a significant role, but can
be overlooked if the overall performance is good.
We assume that there might also be some issues with the gain measurement
for the camera but, have not yet explored this further. 

Unfortunately, we had a large number of dropped frames over the course
of our observations, where we lost as many as 10 frames in approximately
ten seconds of exposures.
The camera never achieved the expected frame rate, however it is
difficult to quantify this given the precision of the time stamps.
The software does not have the ability to set a time frame over which to
take images, only a start and stop function.
These factors make it difficult to know how much data is being taken,
especially if frames are dropped during observations.

\subsection{Comparison with SHWFS}

In comparing the dOTF phase reconstructions with the data from the
SHWFS, we now face the following complications: not enough signal
in the dOTF output -making the reconstruction difficult to trust and
therefore hard to compare- and the best signal (Antares) is a binary
system.

\begin{figure}
	\minipage{0.5\textwidth}
	\includegraphics[width=\linewidth]{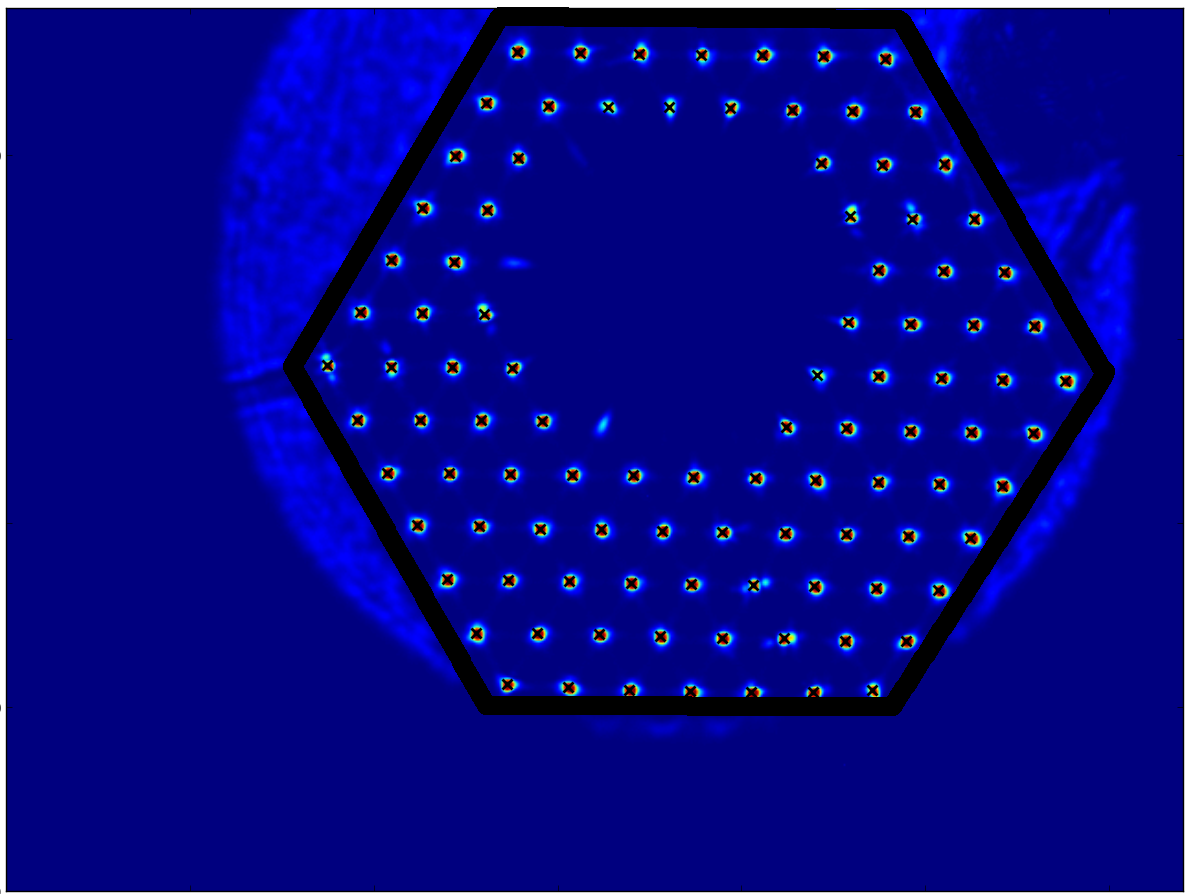}	
	\endminipage\hfill
	\minipage{0.5\textwidth}
    \includegraphics[width=\linewidth]{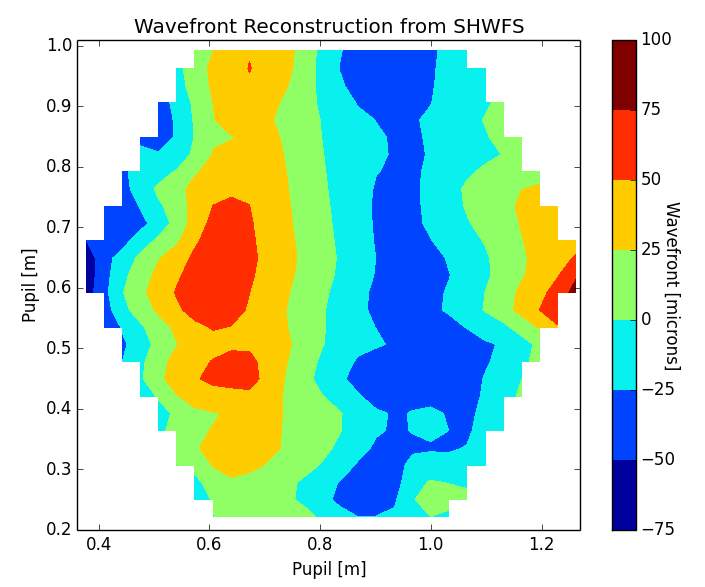}
    \endminipage\hfill
    \caption{\textit{Left:} A pupil plane image taken of Antares with the SHWFS. The hexagon indicates the edges of the capabilities of the SHWFS. Any information on the incoming wavefront outside of this hexagon cannot be reconstructed. For orientation, the polarization obscuration can be seen in the upper right part of the pupil. \textit{Right:} A single frame showing the SHWFS reconstruction of the wavefront for Antares.}
    \label{fig:shwfs}
\end{figure}

Figure~\ref{fig:shwfs} shows one frame of Antares taken by the SHWFS on
the left and the reconstructed wavefront on the right.
The hexagon we see marks the edges of the area in the pupil that the
SHWFS can reconstruct.
It is only within this hexagonal area that we can compare our data.
We know that the comparison of these data is limited by a time and
spatial offset.
It is seen in figure on the left in Figure~\ref{fig:shwfs} that the
orientation of the pupil for the SHWFS is different than what we see
with our polarization camera (this is obvious by the location of the
polarization finger in the upper right of this figure if compared to
Figure~\ref{fig:poltest_finger}).
In the figure on the right in Figure~\ref{fig:shwfs} we give an example
of the wavefront reconstruction from the SHWFS for Antares. 

As for the time offset, we know that the images taken by the
Shack-Hartmann system were taken at around the same time as the
polarization camera images, however since the time stamps for both data
are precise only to within one second (while the exposures are on the order of
milliseconds), we could not determine a more precise offset.
Lastly, given the performance of the polarization camera, we can assume
that there are some significant time gaps in the dOTF data which we
could not recover from our observing logs.
Due to these complications, we are unable to directly compare 
the outputs of the SHWFS and the polarization dOTF WFS.

\section{Conclusions}

We show that using a polarization finger to provide pupil
diversity works and that we can detect a change in the phase of the
wavefront.
However, we are unable to confirm that the dOTF is accurately 
reconstructing the incoming wavefront at the telescope.
The two reasons for this are: (i) the lower than expected signal to noise of the camera
on bright stars and (ii) intermittent and unexplained frame drops from
the polarization camera preventing wavefront synchronization between the
two wavefront sensors.
Future work includes investigating the polarization camera drivers to
make them more robust and optimizing the optical design to increase the
throughput of the instrument.

\acknowledgments 
 
This paper uses observations made at the South African Astronomical
Observatory (SAAO).
We thank the NWO and NRF for support of the project.
We would like to thank the mechanics and the workshop at the SAAO for
making the instrument mounting plates on very short notice. 


\bibliography{brooks_spie_9912-02}
\bibliographystyle{spiebib} 

\end{document}